\documentclass[times,twocolumn,final]{elsarticle}
\usepackage[switch]{lineno}
\usepackage{medima}
\usepackage{framed,multirow}

\usepackage{amssymb}
\usepackage{latexsym}

\usepackage{url}
\usepackage{xcolor}

\usepackage{hyperref}

\usepackage[numbers]{natbib}
\usepackage[caption=false,font=normalsize,labelfont=sf,textfont=sf]{subfig}
\usepackage{multicol}
\usepackage{enumitem}
\usepackage{amsmath}
\usepackage{float}

\usepackage{algorithm}
\usepackage{algpseudocode}

\definecolor{newcolor}{rgb}{.8,.349,.1}

\begin{document}

\verso{Rongjun Ge \textit{et~al.}}

\begin{frontmatter}

\title{VCC-DSA: A Novel Vascular Consistency Constrained DSA Imaging Model for Motion Artifact Suppression}

\author[1]{Rongjun \snm{Ge} \fnref{co-first}}
\author[2]{Weilong \snm{Mao} \fnref{co-first}}

\author[3]{Jian \snm{Lu} }
\author[3]{Rong \snm{Yan} }
\author[4,5,6]{Yikun \snm{Zhang} }
\author[4,5,6]{Peng \snm{Yuan}}
\author[2]{Jun \snm{Xiang}}
\author[4,5,6]{Hui \snm{Tang}}
\author[4,5,6]{Guanyu \snm{Yang}}
\author[4,5,6]{Yudong \snm{Zhang}}
\author[4,5,6]{Yang \snm{Chen}\corref{cor}}
\ead{chenyang.list@seu.edu.cn}
\author[7]{Shuo \snm{Li}}
\cortext[cor]{Corresponding author} 
\fntext[co-first]{These authors contributed equally to this work.}

\address[1]{School of Instrument Science and Engineering, Southeast University, No.2 Sipailou, Nanjing, Jiangsu 210096, China}
\address[2]{X-Ray Department, United Imaging Healthcare Limited Company, Shanghai, China}
\address[3]{Interventional and vascular surgery, Zhongda Hospital Affiliated to Southeast University, Nanjing, China}
\address[4]{Jiangsu Provincial Joint International Research Laboratory of Medical Information Processing, Laboratory of Image Science and Technology, Southeast University, No.2 Sipailou, Nanjing, Jiangsu 210096, China}
\address[5]{Key Laboratory of New Generation Artificial Intelligence Technology and Its Interdisciplinary Applications (Ministry of Education), Southeast University, No.2 Sipailou, Nanjing, Jiangsu 210096, China}
\address[6]{School of Computer Science and Engineering, Southeast University, No.2 Sipailou, Nanjing, Jiangsu 210096, China}
\address[7]{Department of Biomedical Engineering, Case Western Reserve University, Cleveland, USA}

\received{xx xx 2024}
\finalform{xx xx 2024}
\accepted{xx xx 2024}
\availableonline{xx xx 2024}

\begin{abstract}

Digital Subtraction Angiography (DSA) is a clinically significant imaging technique for diagnosing cerebrovascular disease, as gold-standard.
However, the artifacts caused by motion of high-attenuation tissues such as bones, teeth, and catheters, seriously reduce the visibility of blood vessels. 
Currently, there are two types of methods for DSA: registration-based methods for motion compensation, and learning-based DSA synthesis from live images.
However, the performance of motion compensation is limited by the accuracy of registration. And the practical application of learning-based DSA synthesis is hindered by the ill-posedness of the input paradigm and the high matching requirements of data, causing degradation in fake and broke structure.
This paper presents a novel Vascular Consistency Constrained DSA Imaging Model  (VCC-DSA) for robust motion suppression and precise vascular imaging with the following designs: 
1) We specially design a Learning-based Subtraction Mapping Paradigm, so that the ill-posed problem of existing learning-based methods can be solved to enhance the stability of the algorithm.
{2) Our model effectively develops Residual Dense Blocks and details-shortcut to improve the performance under complex structures, such as moving bones overlapping with blood vessels, and small features, like peripheral vessels.}
{3) An innovative Vascular Consistency Strategy is proposed to extract intrinsically consistency from the various relative motions in mask-live images, so that spontaneously
distils the vascular structure with contrast-agent development and robustly suppress motion artifacts, and also naturally alleviates the high matching requirements of data.}
4) We creatively design a Mixup-based Data Self-evolution Strategy for data-intra self-enhancement in training loop, so that the training data gains dynamically optimized to promote model better learning the vascular features, and excluding the irrelevant structures in live/mask image and even the inevitable-artifacts/fake-structure in label.
Prospectively, to further evaluate practical value, an actual general anesthesia animal experiment is specially conducted, besides the assessment on human clinical data. 
Compared with other method, our model improves the PSNR and SSIM by 73.4\% and 8.56\%, respectively.


\end{abstract}


\end{frontmatter}



\begin{figure*}[t]
    \centering
    \resizebox{\textwidth}{!}
    {\includegraphics{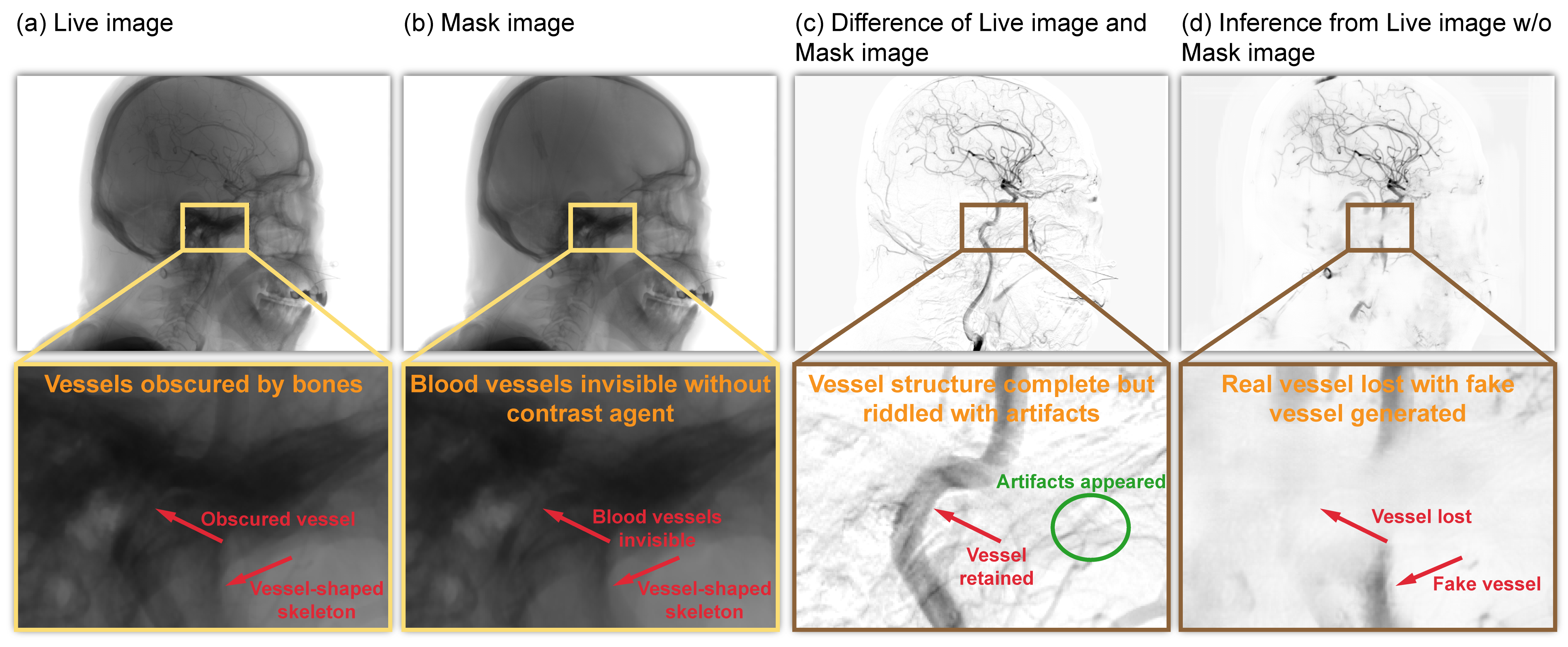}}
    \vspace{-0.2in}
    \caption{The challenges in the DSA task. (a) Live image, where the blood vessels in the low-contrast region are difficult to distinguish.
    (b) Mask image, which does not contain blood vessels.
     (c) The subtraction result of the live image and the mask image retains all the vascular, but there are many motion artifacts. (d) The result inferred from the live image without the mask image, the blood vessels in the overlapping region are difficult to identify.}\label{fig3}
\end{figure*}

\section{Introduction}
\label{sec1}

Digital Subtraction Angiography (DSA) is considered as the gold standard for diagnosing acute cerebrovascular diseases (\cite{arias2015utility, assarzadegan2021evaluation,shaban2022digital,chappell2003comparison}), thanks to it excels in displaying vascular conditions. 
So that precise vascular imaging with motion robustness in DSA is significant and urgent for clinical practice.
However, the impact of patient intrinsic motion before and after contrast agent injection still hinders imaging performance.        
Currently, widely-used time subtraction technology eliminates background information by subtracting the mask image before the injection of the contrast agent from the live image of the filled vessel, as shown in Fig.~\ref{fig3}(a)-(c).
However as can been seen, the motion artifacts inherently existing in time differences dramatically affect vascular observation,
which is still an intrinsic problem due to the actually unavoidable physiological motion of the patient. 
Ideally, if both the live and the mask images perfectly match in structure and grayscale, a blood vessel-only image is gained for doctors to readily diagnose and perform treatments such as endovascular embolization and arterial stenting.
However, when there are disparities in structure and grayscale inevitable between the live and the mask images, {significant residual artifacts} appear on the background or even overlap with the blood vessels. It seriously affects the diagnosis and localization of lesions such as thrombosis and endovascular aneurysm by doctors, as shown in Fig.~\ref{fig3}(c).
{During the angiography procedures of interventional surgery, motion disparity between mask images and live images is inevitable, primarily stemming from involuntary patient motion caused by inherent physiological responses. This phenomenon arises from autonomic reactions to external stimuli (such as cardiac discomfort induced by contrast agent administration  (\cite{seyferth1983efficacy})) and spontaneous muscle contractions (\cite{schafer1886rhythm}), which collectively create a physiological feedback loop that exacerbates image misalignment. The inherent nature of these unconscious bodily responses fundamentally limits synchronization between pre-contrast images and subsequent dynamic acquisitions. Motion disparity thus brings differences in structure and grayscale inevitable between the live and the mask images, causing messy artifacts and loss/coverage of vessel in DSA imaging.}

Traditionally, preliminary registration algorithms are used to reduce motion artifacts between the mask image and the live image firstly before digital elimination (called as ``subtraction") of them.
However, the performance of such motion compensation is limited and fully dependent on the accuracy of registration algorithms which is still open and challenging task, due to the uncontrollability, complexity and multi-dimensionality of the patient's unpredictable movement.
Therefore, due to these inevitably unhandled motions, subtracting the mask image from the live image has great risk to remarkably bring serious structural differences for vascular imaging, especially in critical region even with small motion.
Additionally, the grayscale of the vascular structure and the motion artifacts are structurally similar and spatially overlapping, making it difficult for filtering methods to further process the residual artifacts.
Therefore, in the face of body motion {that cannot be} fully eliminated in traditional registration-based methods, directly mining the mapping relationship between the mask/live image and the vascular imaging is a great alternative.

The current learning-based methods attempt to solve the motion artifacts by learning the mapping from the live image to the DSA image.
However, these methods are hindered in their application in DSA due to the strict requirements for artifact-free DSA image as learning target.
As a dense regression task for pixel-level extraction of different physiological structures, the motion difference between the mask image and the live image cannot be completely eliminated by post-processing once it occurs.
Furthermore, due to the unavoidable physiological motion, the paired data without relative motion is too rare and almost unobtainable. 
The above problems, coupled with the unlabelability of the data, limit the application of these learning-based methods in DSA.

Some learning-based works (\cite{gao2019deep, ueda2021deep, yonezawa2022maskless}) attempt to use synthetic algorithms to abstract the DSA problem into a style transfer problem from live images to DSA images.
However, the similarity between the skeleton and the contrast agent in structure and attenuation coefficients makes it an ill-posed problem to extract only the vascular structure from the live image, causing the fake vessel and the lost as shown in Fig.~\ref{fig3}(d).
These algorithms mistake non-vascular structures as blood vessels, such as the skeleton which is similar to the blood vessels in the live image.
On the one hand, compared with motion artifacts, the fake vascular structure is more deceptive for clinical because synthetic algorithms convert it into a morphology that is very similar to the blood vessel.
On the other hand, the presence of the skeletal structure often causes the contrast agent to the overlap of bone as shown in Fig.~\ref{fig3}(a). This overlap makes it challenging to distinguish the bone from the live image and extract the vascular structure. Therefore, the method of inferring directly from the live image often fails when dealing with complex data with similar overlapping structures.
Especially, extracting small structures such as peripheral vessels from complex and similar backgrounds also increases the difficulty for these works.
Clearly, the above ill-posedness is arised from the absence of the background information to perceive contrast agent flow, i.e., subtracted image for DSA, to reconstruct vascular imaging.
These methods are thus hindered for practical applications.

\begin{figure*}[t]
    \centering
    \resizebox{0.73\textwidth}{!}
    {\includegraphics{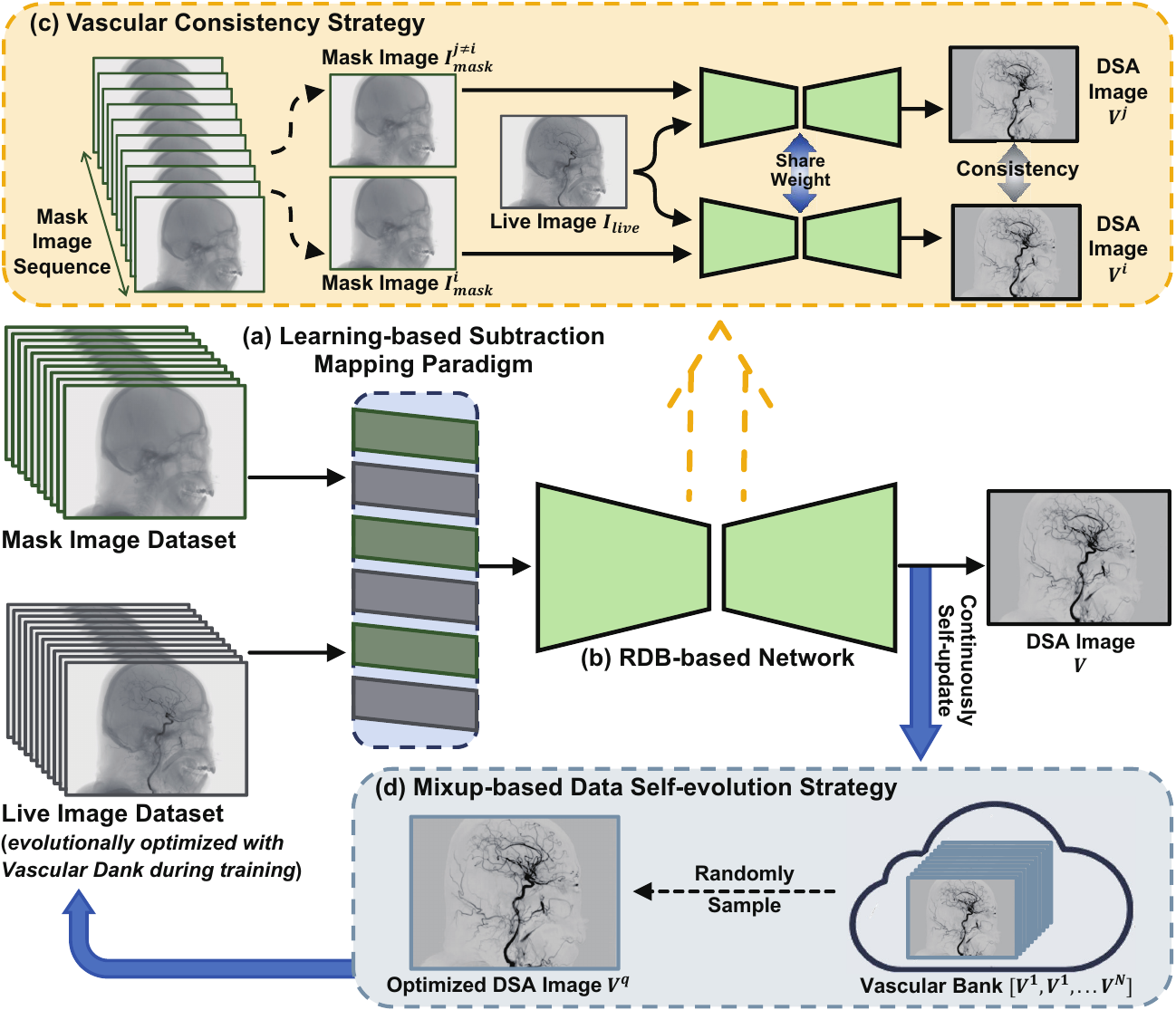}}
    \caption{
        Overall architecture of the proposed VCC-DSA imaging model. It benefits from four special designs: (a) Learning-based Subtraction Mapping Paradigm for solving ill-posed problem and stable DSA imaging, {(b) Residual Dense Blocks-based network for detail enhancement, (c) Vascular Consistency Strateg for motion robustness,} and (d) Mixup-based Data Self-evolution Strategy for vessel boosting.
        }
    \label{fig1}
\end{figure*}

In this paper, to achieve precise vascular imaging for clinical, especially surgery process like thrombus removal, arterial stent implantation, and arterial balloon dilation, and etc.,
we propose a Vascular Consistency Constrained DSA Imaging Model (VCC-DSA). As shown in Fig. \ref{fig1}, it benefits from the following four specially designed elements:
1) A novel Learning-based Subtraction Mapping Paradigm (LSMP) is designed to incorporate background information from typical DSA sequence data to solve the ill-posedness.
The ill-posedness arises from the absence of the background as the subtrahend during the subtraction process, so we provide background information to the algorithm to solve this problem.
Benefiting from the utilization of background information, the ill-posedness is greatly alleviated, so that the ability and stability of vascular recognition are improved when the tissue overlaps and the vascular-like structure appears.
{
2) Residual Dense Blocks (RDB) and details-shortcut are deployed in the network to
 preserve the details of the vascular end.
The vascular structure presents small curved features, and the structure we adopt alleviates the loss of the vascular end during the network propagation.
3) A Vascular Consistency Strategy (VCS) is proposed to
{extract intrinsic consistency from the various relative motions in mask-live image pairs, by penalizing the discrepancy between DSA imaging with different mask images. It thus spontaneously distills the intrinsic vascular structure enhanced by contrast-agent development, robustly suppresses motion artifacts, and naturally alleviates the strictly matching requirements of the data without motion.}
The parameter controllability gives our model strong adaptability, so that the model can control the degree of motion artifacts according to the different motion intensity.
}
4) A Mixup-based Data Self-evolution Strategy (MDSS) is developed to iteratively optimize the vascular structure information in training labels. It leverages the reduction of artifacts and improvement of vascular structure brought about by the self-supervision of consistency strategy. As a result, the network can better learn vascular features and exclude irrelevant structures, further improving the performance.

The contributions of this work are summarized as following:
\begin{itemize}
\item A new VCC-DSA model is designed to achieve the structurally clear DSA imaging for precise vascular extraction with motion artifact suppression, even for the tiny blood vessels, and the inevitable artifacts in learning labels. It further eliminates the severe dependence on artifact-free DSA learning targets, which are difficult acquired and require laboriously manual vascular labeling and painstaking selection of approximate no-motion data in other methods.
\item 
A novel Learning-based Subtraction Mapping Paradigm that incorporates mask images from clinical DSA examination sequence data to provide background information is proposed, so that solve the ill-posed problem and enables the stable DSA imaging results.
{\item
A effective Residual Dense Blocks-based network that performs details-shortcut to transfer the detailed vascular information with thin structure is employed, so that the vascular end in the anatomical structure can be better identified and preserved.}
{\item 
An innovative Vascular Consistency Strategy is proposed to extract intrinsically consistency from the various relative motions in mask-live images, so that spontaneously distils the vascular structure with contrast-agent development and robustly suppress motion artifacts.}
\item A creative Mixup-based Data Self-evolution Strategy is put forward for data-intra self-enhancement in training loop, so that the training data gained dynamically optimized to promote model better learning the vascular features, and excluding the irrelevant structures in live/mask image and even the inevitable-artifacts/fake-structure in label.
\item Prospectively, an actual general anesthesia animal experiment is specially conducted to evaluate practical value of our proposed DSA imaging technology.     
\end{itemize}

\section{RELATED WORK}
Due to the asynchrony of the mask image and the live image acquisition, and the movement of the patient's organs and unavoidable motion (such as heart and respiratory motion), the position of the background tissue around the blood vessel often changes.
Therefore, directly subtracting these images will synthesize motion artifacts in the synthesized DSA image.
Traditional DSA algorithms perform registration through motion compensation to eliminate artifacts. 
Differently, synthesis-based subtraction angiography abstracts the problem into image synthesis from live images to DSA images.
In the following, we will make a comprehensive introduction to the related work of these two types of methods for DSA.

\begin{figure*}[tbp]
    \centering
        \resizebox{0.85\textwidth}{!}
        {\includegraphics{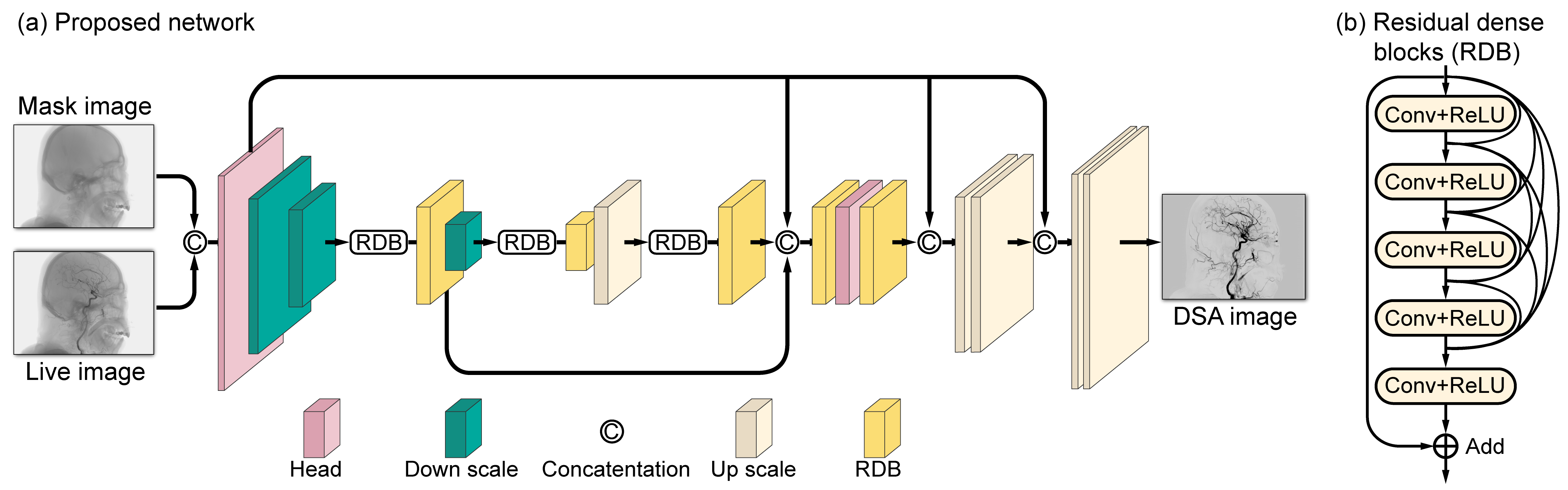}}
        \caption{
            (a) The proposed RDB-based network. The network consists of two parts, namely, the feature extractor and the feature fusion. The feature extractor consists of three downsampling and two RDBs. The feature fusion consists of three upsampling and one RDB. Finally, we use a convolutional layer to map the fused features to the DSA image.
            (b) Residual dense blocks (RDB). The RDB consists of multiple densely concatenated units. The output of the last convolutional layer is added to the input to obtain the final output of the RDB.
            }
        \label{fig2}
\end{figure*}

\subsection{Registration-based Subtraction Angiography}
Traditional registration-based subtraction angiography has been the mainstream method of DSA for a long time.
The development of registration-based subtraction angiography is mainly achieved by improving the accuracy of registration.
In the following, we will introduce the development of registration algorithms related to DSA.

In the 1980s, traditional registration was performed through manual annotation (\cite{harrington1982digital, levin1984digital}).
However, manual annotation can only provide basic image transformations, and is difficult to cope with complex patient movements. While reducing some existing artifacts, it often introduces new artifacts (\cite{levin1984digital}).
To improve local controllability, in 1987, \cite{pickens1987digital} proposed defining the geometric transformation of the mask image using a second-order polynomial, while the matching of control points still relied on manual selection.
In 1999, \cite{meijering1999image} proposed a registration technique based on control points. Their algorithm was the first to be fast enough to be accepted and integrated into clinical applications (\cite{nejati2014multiresolution}).
The more complex algorithms use image features to select control points, including the centerline of the blood vessel (\cite{cao2005dsa}), strong edges in the live or mask image (\cite{taleb1998image, bentoutou2002invariant, bentoutou2005automatic, bentoutou20053, meijering1999retrospective}), curvature (\cite{wang2009iterative}), and structured regions (\cite{nejati2010fast}). Once the corresponding control points of the live and mask images are obtained, the parameters of the registration model can be estimated. This model includes affine transformation (\cite{buzug1996using, buzug1997histogram, buzug1998image}), piecewise linear transformation (\cite{meijering1999image}), and complex transformations such as B-spline transformation (\cite{nejati2013nonrigid}) or thin-plate spline (\cite{cao2005dsa, bentoutou2002invariant}) transformation.
{As can be seen, these above-mentioned methods mainly developed to spatially align the mask image and live image for minimizing the patient motion between before and after contrast agent injection. Based on prior registration-based methods, \cite{R1-1} presented a motion artifact reduction method for DSA sequence images by combining rigid registration with guided filtering, utilizing the latter technique to preserve image contrast during artifact elimination. Developed from the pixel registration before (\cite{taleb1998image, meijering1999image, bentoutou2005automatic}), \cite{R1-5} designed a stratified registration approach by using quad-trees to generate the non-uniform grid of control points and obtaining the sub-pixel displacement offsets using Random Walker, to improve DSA image quality. Instead of the previous control points extraction and registration methods (\cite{meijering1999retrospective, bentoutou2002invariant, wang2009iterative, nejati2014multiresolution}), \cite{R1-10} developed to learn the registration deformation field between the output contrast-removed image and the pre-contrast image to correct the motion. And \cite{R1-11} combined HyperMorph (\cite{R1-12}) with the vessel similarity loss to optimize background registration.}

\textbf{Discussion:}
Actually, the motion artifacts in two-dimensional projection imaging of DSA is caused by patient's instinctively three-dimensional activity, not just movement within the plane.
Thus, although registration-based subtraction angiography have made progress under the development of registration over the past few decades, the problem of motion artifacts has not been fully resolved through these two-dimensional registration methods due to the complexity and multidimensionality of patient movement.
The patient's movement is mainly manifested as three-dimensional rotation and translation. Due to the patient's semi-transparent nature under X-rays, this movement cannot be matched by two-dimensional registration methods after being projected onto a two-dimensional plane.
Given the shortcomings of two-dimensional registration under multidimensional motion, we propose a new learning-based DSA algorithm that does not rely on the above explicit registration.
Instead, we use a consistency strategy to learn richer feature information without using the two-dimensional registration result as a registration medium, and implicitly align the motion differences between the mask image and the live image, for structure preservation and artifact suppression.

\subsection{Learning-based DSA Synthesis}
Some studies (\cite{gao2019deep, ueda2021deep, yonezawa2022maskless}) abstract the DSA into a style transfer task by synthesizing DSA images from live images. 
\cite{gao2019deep} first proposed a synthesis-based subtraction angiography to synthesize DSA images from live images. This method constrains the generator by adding an additional discriminator so that the synthetized result by the generator are closer to the DSA images visually.
Subsequently, \cite{ueda2021deep} and \cite{yonezawa2022maskless} made similar attempts with more data using the pix2pix network (\cite{isola2018imagetoimage}).
To improve the performance of the algorithm in low-contrast regions, \cite{kimura2020virtual} used different networks for different regions to enhance the accuracy of blood vessel extraction in each specific region.
In recent studies, it has been observed that the degree of artifacts varies across different anatomical regions. To address the issue of poor data quality in regions with severe artifacts, \cite{crabb2023deep} employed transfer learning as a means to mitigate the problem.

\textbf{Discussion:}
These methods rely on the quality and quantity of training target artifact-free DSA image. 
However, due to the inevitable issues like heartbeats and involuntary muscle contractions, it is rare to find such clean data in actual clinical settings, making artifact-free DSA collection difficult.
Besides, as shown in Fig.~\ref{fig3}(a) and (b), due to the complexity of the human skeletal structure, there are some skeletal structures that are similar to the blood vessels in terms of attenuation coefficients and structure, making it an ill-posed problem to distinguish the skeleton from the live image and extract the vascular structure.
This ill-posedness leads to results in Fig.~\ref{fig3}(d), where the algorithm incorrectly identifies the skeleton as a blood vessel and erases the real vascular structure, making the method of directly synthesizing DSA images from live images difficult to apply in actual clinical settings.
Since the mask image provides background information about the live image, it can greatly alleviate the ill-posed problem.
However, due to the unresolved motion disparity, it has become the bottleneck of the current method to utilize background information to solve the ill-posed problem, resulting in the instability of the current method.
In our work, we ingeniously addressed the issue arising from the motion disparity between two entities with newly designed consistency strategy. This strategy integrates mask information to tackle the ill-posed problem effectively.

\section{METHODOLOGY}
\subsection{Learning-based Subtraction Mapping Paradigm}
Existing DSA synthesis algorithms take the live image as input, and makes style transfer to synthesise a DSA image directly from it. 
The synthesis process can be described by the following equation:
\begin{equation}
    \begin{aligned}
    V&= \mathcal{S}(I_{live}) \\
    &= \mathcal{S}(B+V)
    \end{aligned}\label{eqwoIP}
\end{equation}
where $I_{live}$ denotes the live image, $V$ denotes {DSA image that solely includes vascular information without motion artifacts}, $B$ denotes the background in the live image, and $\mathcal{S}$ denotes the synthesis algorithm, which is used to transfer $I_{live}$ into DSA image that contains the vascular structure from the live image.
This method abandons the use of background information in the mask image to alleviate the motion problem between the mask image and the live image.

According to the Lambert-Beer theory (\cite{lambert1760photometria}), the live image can be described as the sum attenuation coefficient of the background ${B}$ and the vascular structure $V$.
Therefore, the above synthesis process can be described as an ill-posed process of extracting the addend from the sum.
Due to the complexity of the human skeletal structure, the ill-posedness caused by the live image alone makes it difficult for the algorithm to distinguish between blood vessels and bones accurately.
The ill-posedness is manifested in three main issues in the image: 1) the skeleton is identified and synthesized as a deceptive fake blood vessel; 2) the blood vessel and the skeleton overlap, making the blood vessel ignored; 3) a large area of blur is synthesized, as shown in Fig.~\ref{fig3}(d).

To address these issues and drive precise DSA reconstruction with vascular enhancement and artifact suppression,
we propose a novel Learning-based Subtraction Mapping Paradigm (shown in Fig. \ref{fig1} (a)) that incorporates mask image sequence to provide background information, so that the ill-posed problem can be solved and the stable results can be obtained. Our input paradigm can be described as: 
\begin{equation}
    \begin{aligned}
    V &= \mathcal{F}(I_{mask},I_{live})\\
      &= \mathcal{F}(\mathcal{W}(B,R),I_{live})\\
      &= \mathcal{F}(\mathcal{W}(B,R),B+V)\label{eqwIP}
    \end{aligned}
\end{equation}
where $\mathcal{F}(\cdot)$ symbolizes our learning-based DSA reconstruction network. 
{To extract and preserve the vascular structure under complex structures, it is designed with details-shortcut, Residual Dense Blocks, and successive convolutional feature extraction and fusion, as well as downsampling and upsampling, as shown Fig.~\ref{fig2}. The details-shortcut is developed to connect Head block of shallow layer with multiply deep decoding layers to retain the detailed features of the blood vessels, while the Residual Dense Blocks with densely connections extract and fuse multi-scale structural features. The convolutions adopt $3\times3$ kernel, with $32\to64\to128\to128$ channels in feature extraction and reversed in fusion.}
$I_{mask}$ is the mask image that has relative motion with the live image $I_{live}$, and $B$ means the decoupled background information.
$R$ represents the random motion fields of the patient's unconscious movement, and $\mathcal{W}(\cdot)$ is the warp operation.
$V$ indicates the DSA image that solely includes vascular information without motion artifacts. 
{Specifically, with the Learning-based Subtraction Mapping Paradigm, the reconstruction network takes the combinations of mask image $I_{mask}$ and live image $I_{live}$ as input, to perceive and decouple the background $B$ after contrast agent injection for clear vascular $V$ imaging without artifacts. During each training iteration, for each $I_{live}$ from live image dataset, the mask image dataset randomly provides $I_{mask}^i$ and $I_{mask}^{j\ne i}$ acquired at different time points $i$ and $j$ before the injection of the contrast agent. Then, the combinations $\{I_{mask}^i, I_{live}\}$ and $\{I_{mask}^{j\ne i}, I_{live}\}$ are input into the learning-based DSA reconstruction network respectively, with shared weight. Subsequently, the Vascular Consistency Strategy constraints the network training to extract the consistence, i.e. vascular imaging, in these different relative motion combinations $\{I_{mask}^i, I_{live}\}$ and $\{I_{mask}^{j\ne i}, I_{live}\}$. During inference, the there is no need for two input combinations. Each live image only needs to be combined with any one mask image from the same patient and input into the network to obtain the DSA imaging with motion artifact suppression and precise vascular details.}

\subsection{RDB-based Network for Detail Enhancement}

{DSA sequence has} following two characteristics: 1) The structure of blood vessels is very small, so the network needs to have sufficient detail feature extraction ability to identify the blood vessels; 2) The complexity of the artifacts requires the network to have a sufficient receptive field to identify the artifacts. 
According to these characteristics, the details-shortcut is achieved by downsampling the output of the Head block and cascading it with the output of each upsample in decoding to retain the detailed features of the blood vessels.
And RDB is built with multiple densely connected convolutional layers at bottom of the network. So that it can effectively extract the global features of the image while retaining the detailed features of the image.

As detailed in Fig.~\ref{fig2}(a), the network consists of three downsampling and three upsampling operations.
After two downsampling operations, we extract features through the RDB.
And each RDB comprises five convolutional layers that are densely connected by residual connections, where the output of each convolutional layer is concatenated with the output of all the previous convolutional layers, as shown in Fig.~\ref{fig2}(b). The output of the last convolutional layer in each RDB is added to the input to obtain the final output of the RDB.
Additionally, the Head block is a convolutional layer with an input channel of 2 and an output channel of 32, while the corresponding Tail block has an input channel of 64 and an output channel of 1. The kernel size for all these layers is set to $3\times3$. The filter numbers for the three downsampling and upsampling operations are 64, 128, and 128, respectively.
To preserve the vascular details, we specifically downsample the output of the first Head block and the first RDB, and concatenate it with multiple upsampling modules to retain the vascular information. {The computational complexity of the network is 96.56G FLOPs, with 19.32M parameters.}

\begin{figure}[tbp]
    \centering
        \resizebox{0.4\textwidth}{!}
        {\includegraphics{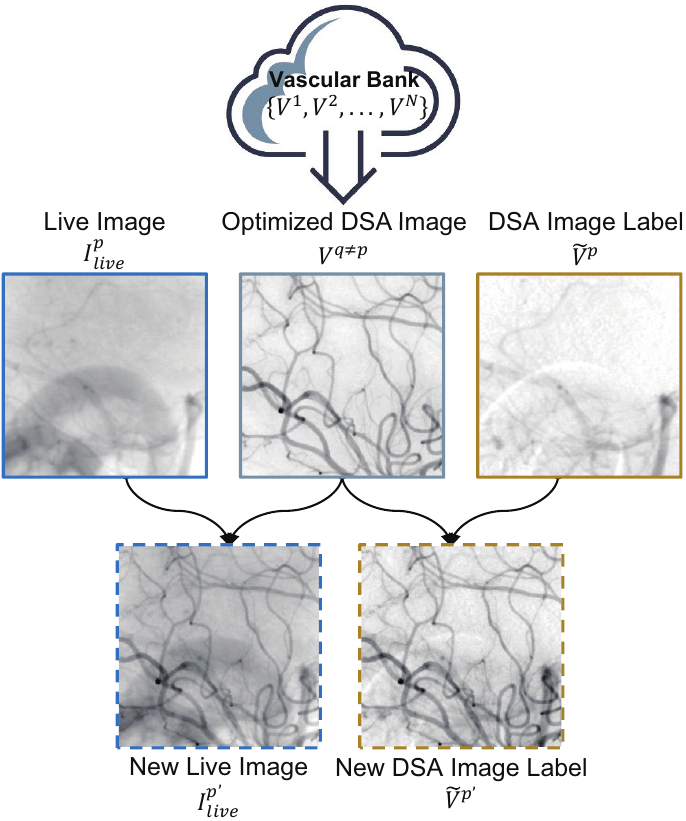}}
        \vspace{-0.1in}
        \caption{
MDSS evolutionally optimizes training data for new live image with more complex structures and label with more accurate vascular information to enhance model understanding on blood vessels, by randomly sampling the optimized DSA image from Vascular Bank to mixed up onto the original leaning label and the corresponding live image in training. 
            }
        \label{fig_MDSS}
\end{figure}

\subsection{Vascular Consistency Strategy for Motion Robustness} 
Although the background information in Eq.\eqref{eqwIP} can alleviate the ill-posed problem, some motion artifacts may still exist since the inherent motion disparity between the mask image and the live image.
As shown in Fig. \ref{fig1}(b), we propose a novel Vascular Consistency Strategy (VCS) to extract intrinsically consistency from the various relative motions in mask-live images, so that the vascular structure with contrast-agent development gets maintained and motion artifacts is robustly suppressed. 
It thus promotes the results-inter structural-constraints for DSA imaging reconstructed from the live image and its moved mask image sequence $\{I_{mask}^i,I_{live}\}$,
and further eliminates the limitation of needing the manual vascular labeling and the painstaking selection of approximate no-motion data in generally supervised training.

The vascular imaging for DSA is invariant to the any motions among mask images before agents injection. It is formulated as: 
\begin{align}
    &\mathcal{F}(I_{mask}^i,I_{live}) = \mathcal{F}(I_{mask}^{j\neq i},I_{live})  \nonumber
\\ \Rightarrow \ &\mathcal{F}(\mathcal{W}(B,R^i),I_{live}) = \mathcal{F}(\mathcal{W}(B,R^{j\neq i}),I_{live})\label{eq2}
\end{align}
where $I_{mask}^i$ and $I_{mask}^{j\neq i}$ are {mask images} at different time points, with different motion fields $R^i$ and $R^{j\neq i}$.
So that the proposed VCS is able to optimize model for structure imaging and artifacts supression, as:
\begin{equation}
    \underset{\mathcal{F}}{min}\ \mathcal{L}_{con} = \mathcal{F}(\mathcal{W}(B,R^i),I_{live}) - \mathcal{F}(\mathcal{W}(B,R^{j\neq i}),I_{live}) \label{eq3}
\end{equation}

VCS thus drives DSA imaging model focus on consistent structure, i.e. vascular, so that the impacts of relative motion between mask images and live images, and the artifacts in the training labels are eliminated.
In collecting clinical data for DSA examination, there often acquires two original parts: 1) The first part consists of a series of images taken without the contrast agents injection as mask image sequence; 2) The second part contains images with contrast agents, i.e. live image.
Following Eq.\eqref{eq3}, any two mask images are randomly sampled with the corresponding same live image to guide the network on consistent vascular learning and eliminating motion differences. 
Since the two selected mask images are from different time points in the same sequence, there inherently exist relative motion between them. The motion artifacts is suppressed by minimizing the discrepancy between the reconstructed DSA images with different mask images.

Specifically, two synchronous forward propagations are deployed during training, as shown in Fig.~\ref{fig1}(a).
With different mask-live image combinations $\{I_{mask}^i, I_{live}\}$ and $\{I_{mask}^{j\neq i}, I_{live}\}$, the paired consistent DSA results $V^i$ and $V^j$ are gained.   
The discrepancy between $V^i$ and $V^j$ is caused by the relative motion between the different mask images $I_{mask}^i=\mathcal{W}(B,R^i)$ and $I_{mask}^j=\mathcal{W}(B,R^j)$, as sharing the same live image $I_{live}$ with same vascular imaging. According to Eq.\eqref{eq3}, we minimize the difference between these two DSA results to update model in backward propagation.

Constrained with the VCS, the model loss function $\mathcal{L}_{total}$ is developed into two parts: 1) The fidelity term $\mathcal{L}_{fid1}$ and $\mathcal{L}_{fid2}$, which are used to guide the content learning on vascular structure;   
2) The consistency term $\mathcal{L}_{con}$ is used to constrain the consistency between the DSA results from different mask images, ensuring robustness to motions before and after contrast agent injection, and highlighting consistent vascular structure in motions.
Thus, the consistency constrained model loss $\mathcal{L}_{total}$ can be formulated as: 
\begin{equation}
    \begin{aligned}
    &\mathcal{L}_{total} =(1-\lambda) \cdot \underbrace{\frac{\mathcal{L}_{fid1} + \mathcal{L}_{fid2}}{2}}_{\text{fidelity}} \ + \ \lambda\cdot\!\!\!\underbrace{\mathcal{L}_{con}}_{\text{consistency}},\\
    &\text{and}\  \left\{\begin{matrix}
    \mathcal{L}_{fid1} = ||V^i-\tilde{V}||_1\\
    \mathcal{L}_{fid2} = ||V^j-\tilde{V}||_1\\
    \mathcal{L}_{con} = ||V^i- V^j||_1
    \end{matrix}\right. ,
    \label{loss}
    \end{aligned}
\end{equation}
where $\tilde{V}$ is the subtraction DSA image inevitably with motion artifacts, playing the role of weak supervision label for vascular structure reference. 
$\lambda$ means the weight to balance loss terms between fidelity and consistency,  
and $L1$ norm is used in loss calculation.

\subsection{Mixup-based Data Self-evolution Strategy for Vessel Boost}
Our proposed model learns and distinguishes between background information and vascular structure information during the training process, relying on the fidelity term and the consistency term in the loss function Eq.\eqref{loss}.
Our aspiration is to preserve the information pertaining to the vascular structure via the fidelity term, while concurrently mitigating the artifacts engendered by motion disparity through the implementation of the consistency term. 
However, in the actual training process, due to the unavoidable artifacts and fake structure in the DSA learning labels $\tilde{V}$ in the fidelity term, the fidelity term thus misguides the network to retain the artifacts and fake structure introduced from $\tilde{V}$. 
Although this part of the artifacts can be suppressed by increasing the weight $\lambda$ of the consistency term, it also cause the weight of the fidelity term to decrease and lead to a decrease in the learning ability of the vascular morphology. 
{To address this contradiction, we creatively design a Mixup-based Data Self-evolution Strategy (MDSS) to further leverage the data quality enhancements from the consistency strategy, thereby optimizing labels and enriching data to bolster the fidelity term’s ability in learning vascular morphology features.}

Accompanying the model training, MDSS evolutionally optimizes the leaning label $\tilde{V}$ with the ``Vascular Bank" $\left \{ V^1,\ V^2,\ ...\ ,V^N\right \} $ which stores the optimized DSA image results, as shown in Fig.~\ref{fig1}(d). 
Specifically, the randomly sampled image $V^{q\neq p}$ from the Vascular Bank is mixed up onto the original leaning label $\tilde{V}^p$ and the corresponding live image $I_{live}^p$ in training, so that new live image $I_{live}^{p'}$ with more complex structures and label $\tilde{V}^{p'}$ with more accurate vascular information are enabled, as shown in Fig.~\ref{fig_MDSS}.
During the entire training loop, the ``Vascular Bank" is continuously self-updated by the model with the optimized vascular structures.
{Consequently, the MDSS augments the vascular discrimination capability of the fidelity term in the loss function (Eq.\eqref{loss}) for precise DSA imaging. This enhancement is achieved through the superposition of more accurate vascular information, and the increase in the variety of the blood vessels relative to the confounding elements including background interference, imaging artifacts, and non-vascular anatomical imitations.}
The procedure is formulated as:
\begin{align}
    &\min_{\mathcal{F}}(\mathcal{F}(I_{mask}^p, I_{live}^{p'}) -\tilde{V}^{p'} )  \nonumber
\\ \Rightarrow \ &\min_{\mathcal{F}}(\mathcal{F}(I_{mask}^p, I_{live}^p + V^{q\neq p}) -(\tilde{V}^p + V^{q\neq p}))
\label{MDSSeq}
\end{align}
where $I_{live}^p$ represents the live image, $\tilde{V}^p$ and $I_{mask}^p$ represent the corresponding original DSA image label and mask image, $\tilde{V}^{p'}$ and $I_{mask}^{p'}$ are new ones. $V^{q\neq p}$ means the optimized DSA image 
randomly sampled from the vascular dataset, and $\mathcal{F}(\cdot)$ is our proposed DSA imaging model.

\begin{figure*}[!htb]
    \centering
    \resizebox{0.647\textwidth}{!}
    {\includegraphics{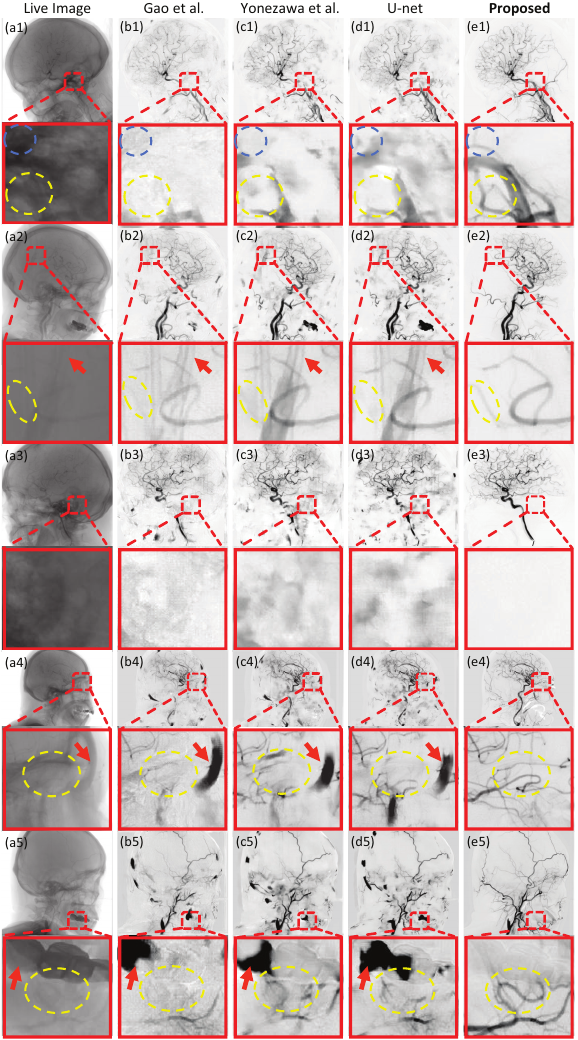}}
    \caption{
        Digital subtraction angiography results.
        Our method (d) has the characteristics of clear vascular structure, no motion artifacts, and clear background, which is conducive to the doctor's diagnosis, while blurred background, vessel loss, and oxygen tube artifacts exist in (a) \cite{gao2019deep}, (b) \cite{yonezawa2022maskless} and (c) U-net (\cite{ronneberger2015u}).
    }
    \label{exp}
\end{figure*}

{\subsection{Algorithm Description}
In summary, the procedure of the proposed can be described as {Algorithm} {\ref{alg}}. In each iteration, given live image $I_{live}$ from the dataset $\mathbb{L}$, the mask image dataset $\mathbb{M}$ provides $I_{mask}^i$ and $I_{mask}^{j\ne i}$ acquired at different time points. Construct inputting combinations $\{I_{mask}^i, I_{live}\}$ and $\{I_{mask}^{j\ne i}, I_{live}\}$, RDB-based Network $\mathcal{F}$ reconstruct DSA images $V^i$ and $V^j$, respectively. With VCS constraints, consistency term $\mathcal{L}_{con}$ is calculated by penalizing the discrepancy between $V^i$ and $V^j$, while fidelity terms $\mathcal{L}_{fid1}$ and $\mathcal{L}_{fid2}$ guide the content learning with original DSA image $\tilde{V}$. The total loss  $\mathcal{L}_{total}$ is gained by combining $\mathcal{L}_{con}$ with $\mathcal{L}_{fid1} + \mathcal{L}_{fid2}$, to guide model update via Adam optimizer. Accompanying the model training, the optimized DSA reconstruction result $V^{q\ne p}$ are further used to update training datasets of live images and DSA images.}

\begin{algorithm}[t]
\small
  \caption{{Training Procedure}}
  \label{alg}
  \begin{algorithmic}[0]
    \Require
      {Live Image Dataset $\mathbb{L}$; Mask Image Dataset $\mathbb{M}$; DSA Image Dataset $\mathbb{V}$}
    \State {$\enspace$ Live image $I_{live} \in \mathbb{L}$; Mask images $I_{mask} \in \mathbb{M}$; DSA image $\tilde{V} \in \mathbb{V}$    }
    \\
    $\bullet $ {{Construct inputting combinations} $\{I_{mask}^i, I_{live}\}$ and $\{I_{mask}^{j\ne i}, I_{live}\}$ }
    \\
    $\bullet $ {{Input $\{I_{mask}^i, I_{live}\}$ and $\{I_{mask}^{j\ne i}, I_{live}\}$ into RDB-based Network $\mathcal{F}$} }
    \State {$\quad \quad \quad  \quad \quad \quad \quad  V^i= \mathcal{F}(I_{mask}^i,I_{live})$}
    \State {$\quad \quad \quad  \quad \quad \quad \quad  V^j= \mathcal{F}(I_{mask}^{j\neq i},I_{live})$}
    \\
    $\bullet $ {{Calculate consistency term (VCS) } }
    \State {$\quad \quad \quad  \quad \quad \quad \quad  \mathcal{L}_{con} = ||V^i- V^j||_1$}
    \\    
    $\bullet $ {{Calculate fidelity term} }
    \State {$\quad \quad \quad  \quad \quad \quad \quad       \mathcal{L}_{fid1} = ||V^i-\tilde{V}||_1$}
    \State {$\quad \quad \quad  \quad \quad \quad \quad      \mathcal{L}_{fid2} = ||V^j-\tilde{V}||_1$}
    \\
    $\bullet $ {{Calculate total loss } }
    \State {$\quad \quad \quad  \quad \quad\mathcal{L}_{total} =(1-\lambda) \cdot \frac{\mathcal{L}_{fid1} + \mathcal{L}_{fid2}}{2} \ + \ \lambda\cdot\!\!\mathcal{L}_{con}$}
    \\
    $\bullet $ {Update the model $\mathcal{F}$ via Adam optimizer}
    \\
    $\bullet $ {{Update live image datase $\mathbb{L}$} }
    \State {$\quad \quad \quad  \quad \quad \quad \quad   I^{p'}_{live}\gets I^{p}_{live}+V^{q\ne p}$}
    \\  
    $\bullet $ {{Update DSA image datase $\mathbb{V}$}} 
    \State {$\quad \quad \quad  \quad \quad \quad \quad   \tilde{V}^{p'}_{live}\gets \tilde{V}^{p}+V^{q\ne p}$    }
    
  \end{algorithmic}
\end{algorithm}

\section{EXPERIMENTS}

\begin{table}[b]
    \caption{
        Quantitative Analysis of our model:
        Compared with the other methods, our model has a remarkable improvement in PSNR, SSIM and inference time.
    }
    \vspace{-0.15in}
    \small
    \setlength\tabcolsep{2.5pt}
    \label{tab1}
    \begin{center}
        \begin{tabular}{cccc}
            \hline \hline
            Method     & SSIM(\%)$\uparrow$ & PSNR($dB$)$\uparrow$  & Inference Time ($s$)$\downarrow$  \\ \hline
            {U-net}
              & 90.502 & 24.01765 &0.086  \\
             \cite{gao2019deep}
              &  91.016 &  24.90746& 0.086  \\ 
            \cite{yonezawa2022maskless}
              &  89.504 & 24.41562& 0.102             \\ 
\textbf{Proposed}  &  \textbf{98.802} & \textbf{43.19122}& \textbf{0.082}     \\ \hline \hline
        \end{tabular}
    \end{center}
\end{table}

\begin{figure*}[!htb]
    \centering
    \resizebox{0.7\textwidth}{!}
    {\includegraphics{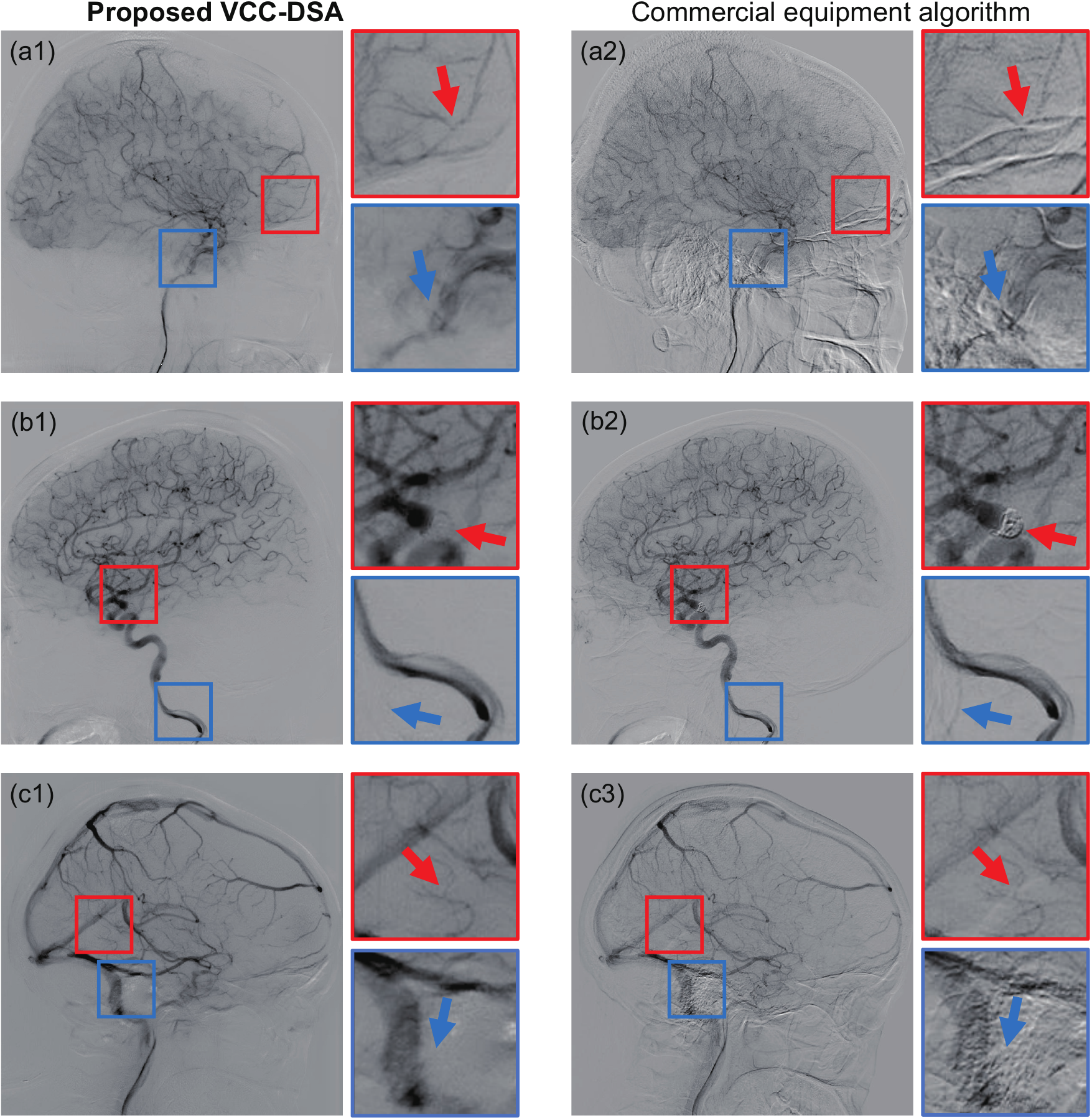}}
    \caption{
        The proposed VCC-DSA method achieves robust cross-dataset generalizability
        }
    \label{figAddition}
\end{figure*}

\subsection{Materials and Configurations}
Our dataset contains two parts, one for human subjects and another is actual general anesthesia animal experiment data.

\begin{itemize}
\item \textbf{Human dataset:} 1) We divided the clinical human dataset collected from United Imaging Healthcare uAngio 960 into training, validaton and test with 781, 39 and 238 images, respectively.
The clinical human test set was used for visual analysis while it is unsuitable for accurate quantitative analysis because artifacts in the DSA image labels will influence quantitative analysis. 
 {2) The dataset acquired from Philips Azurion 7 B20 consists of 1558 imags, and are split into 764:109:685 for training, validaton and test, with no patient overlap. The cross-dataset generalizability can be further evaluated.}
\item \textbf{Actual general anesthesia animal dataset:}  Prospectively, to avoid the influence of artifacts in the labels on the calculation of the indicators during quantitative analysis, we further specially conducted animal experiments for idea DSA image ground truth without motion influence. 
    The animal was tested under general anesthesia to remove the inherent motion characteristics of the subject during the clinical interventional surgery and the subject's non-general anesthesia status often lead to motion artifacts in the DSA image. 
{Thus, the experiment protocol was designed as: the animal was under general anesthesia and received intervention of ventilator; during each DSA acquisition ($\leq 10 s$, , not endangering the life of animal), the ventilator was temporarily paused to achieve an apnea breath-hold, suppressing respiratory motion caused by diaphragm, thoraco-abdomin and the associated coupled movement, and the head was immobilized; the angiographic field of view covered the head/neck, and heart was outside the FOV, so heartbeat did not introduce motion in the imaged region. These measures provide an ideal, motion-free DSA ground truth for quantitative analysis.}
The animal dataset was used in the test phase with human-data trained model, which also reveals the great generality. 
The animal experimental object was a pig with a weight of 50kg and a head thickness of 15cm.
\end{itemize}

\begin{figure*}[!htb]
    \centering
    \resizebox{0.9\textwidth}{!}
    {\includegraphics{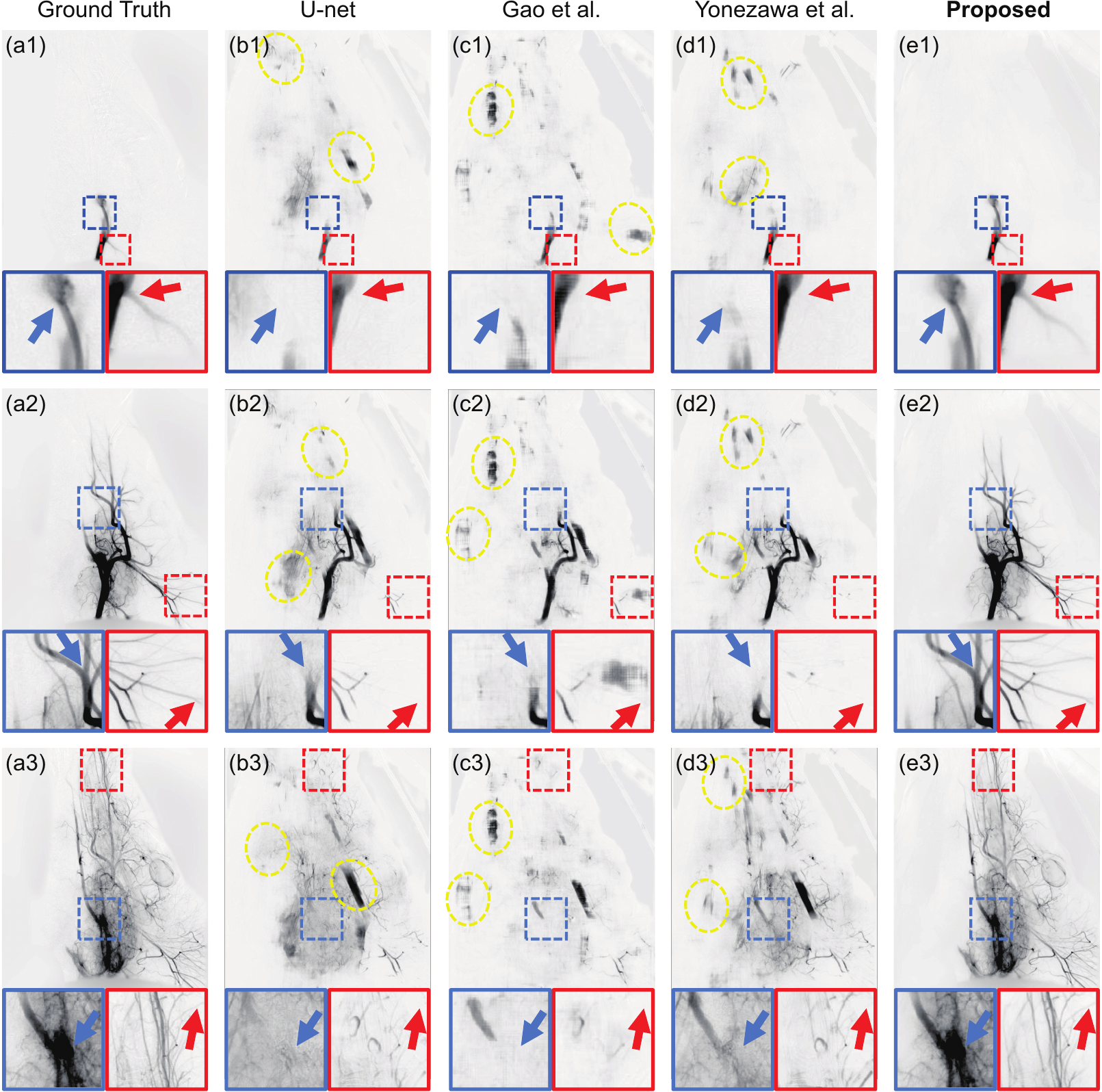}}
    \caption{
{
The proposed method still demonstrates significant DSA imaging superiority with precise vascular imaging and robust artifact suppression, even with cross-species training and testing. 
}
        }
                    \vspace{-0.05in}
    \label{FigAnimal}
\end{figure*}

\begin{figure*}[!t]
    \centering
    \resizebox{0.9\textwidth}{!}
    {\includegraphics{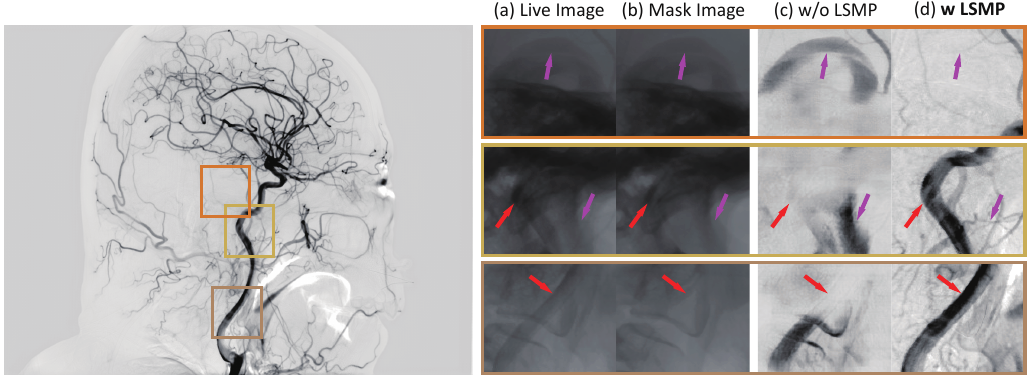}}
    \caption{
        Ablation performance of the Learning-based Subtraction Mapping Paradigm in qualitative analysis.
        Our proposed LSMP (d) with mask image (b) achieves significant advantages in realistic and detailed blood vessel structures. In comparison, the situation of mask image was canceled, the blood vessels in live image (a) are lost in (c) as pointed by the green arrow and the fake blood vessels are generated as pointed by the blue arrow.
        }
    \label{mask}
\end{figure*}

 All data were collected from C-arm cone-beam DSA device. The device provide three parts of data, including: 1) mask image sequence, 2) live image sequence, and 3) DSA image sequence constructed by the device's commercial Pixel Shift algorithm.

Data augmentation is used to expand training data and enhance the model robustness,
with flipping, translation scaling, and rotation.
The images was randomly cropped into size of $256\times256$ to enhance the randomness of the data for robust network training.

In the experiment, the parameter $\lambda$ of the loss function is set as 0.85. 
The Adam optimizer is used with $\beta1=0.9$ and $\beta2=0.999$. The learning rate is $0.0002$. The GPU used for training is a 3080ti with data batch size of 16.

\subsection{Evaluation Metrics}
To quantitatively evaluate the performance of artifact removal and vascular structure preservation, Structural Similarity Index (SSIM) and Peak Signal-to-Noise Ratio (PSNR) are adopted.
 SSIM is defined as:
\begin{equation}
    SSIM(x,y) = \frac{(2\mu_x\mu_y + C_1)(2\sigma_{xy} + C_2)}{(\mu_x^2 + \mu_y^2 + C_1)(\sigma_x^2 + \sigma_y^2 + C_2)} \times 100\%
\end{equation}
where $\mu_x$ and $\mu_y$ present the corresponding mean value, $\sigma_x^2$ and $\sigma_y^2$ are the variances, $\sigma_{xy}$ denotes the covariance, and $C_1$ and $C_2$ are two constants used to stabilize the denominator.

And PSNR is calculated as:
\begin{equation}
    PSNR = 20 * \log10(MAX_I) - 10 * \log10(MSE)
\end{equation}
where $MAX_I$ is the maximum possible pixel value of the image, and $MSE$ is the Mean Square Error.

\begin{figure}
    \centering
    \resizebox{0.48\textwidth}{!}
    {\includegraphics{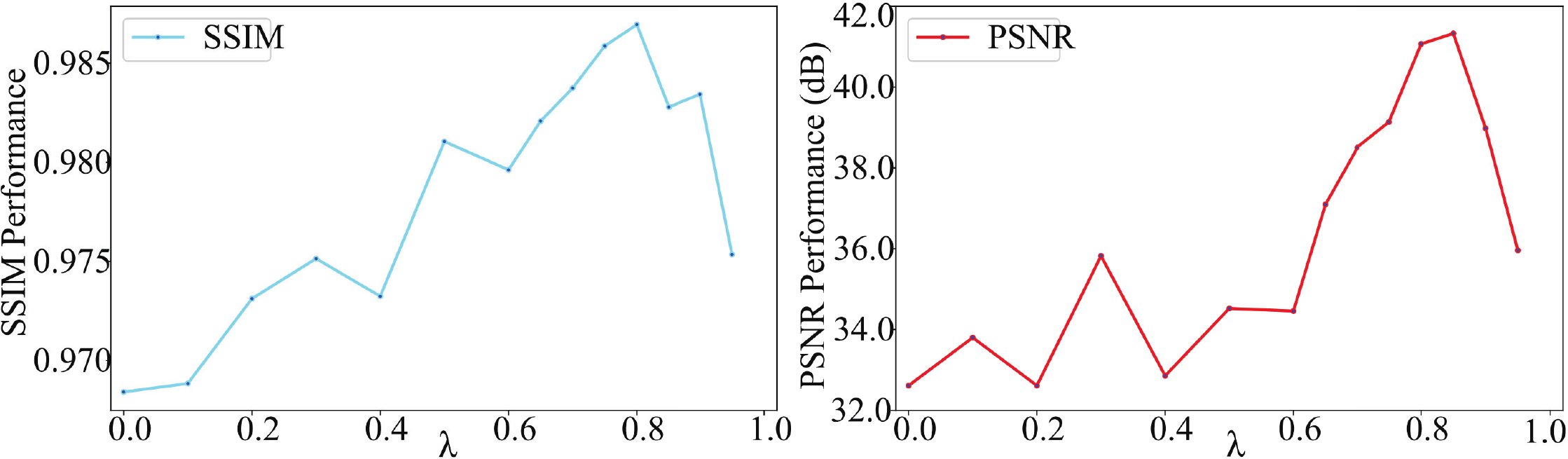}}
    \vspace{-0.3in}
    \caption{
    Ablation performance of Vascular Consistency Strategy in quantitative analysis.
    By gradually increasing the value of $\lambda$, the weight of the consistency strategy is adjusted. Both PSNR and SSIM increase with the increase of $\lambda$. When $\lambda$ is greater than 0.85, PSNR and SSIM decrease.
    }
    \label{fig4}
\end{figure}

\begin{figure*}
        \centering
        \resizebox{\textwidth}{!}
        {\includegraphics{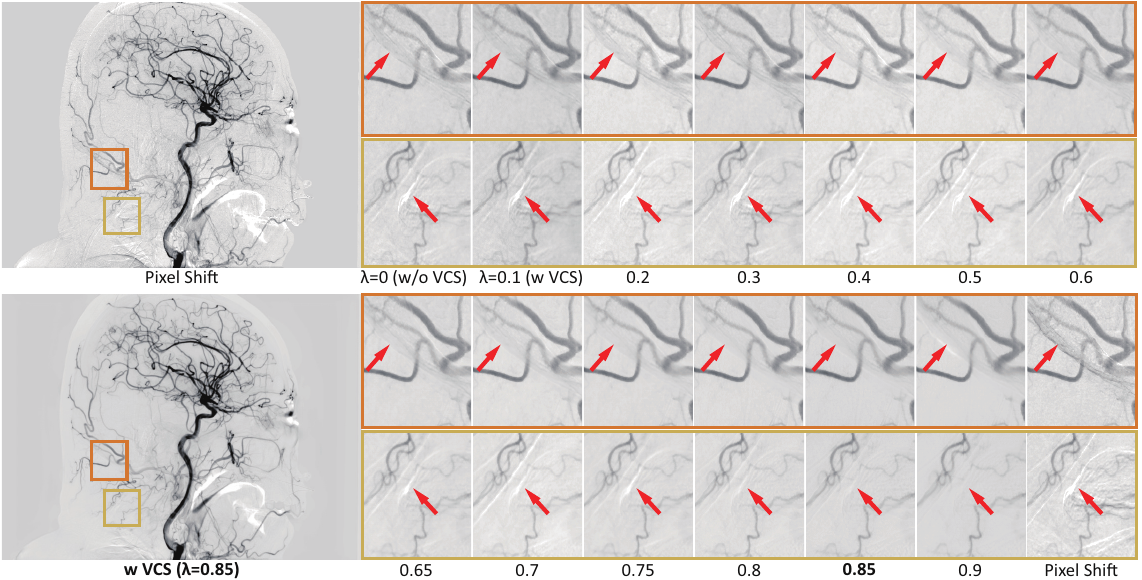}}
        \vspace{-0.29in}
        \caption{
            Ablation performance of Vascular Consistency Strategy in visual comparison experiments with $\lambda$ changes and the Pixel Shift method used in the commercial software of United Imaging Healthcare.
            By gradually increasing the value of $\lambda$, the weight of the consistency strategy is adjusted. We can see that the artifacts gradually decrease and the visibility of the blood vessels gradually increases.
            Compared with the Pixel Shift method, our method can have comparable vascular details and significantly less artifacts.
        }
        \label{lambda}
\end{figure*}

\subsection{Results and Analysis}
In the performance evaluation, clinical human data is used for visual comparison experiments, and animal experimental data is employed for quantitative analysis.

1) In order to demonstrate the effect of the model in real clinical scenario, the clinical human data is used for the visual analysis. The details of artifact removal and vascular structure preservation are explicitly and directly demonstrated with actual visual imaging effect.
 But it is not suitable for quantitative metric calculations, because patients in interventional surgery are rarely under general anesthesia, resulting in almost all acquired clinical DSA image having varying degrees of artifacts.
Using these DSA image with motion artifacts as the ground truth will cause the evaluation metrics just scoring the results on the similar artifacts, which is not in line with the purpose of artifact reduction evaluation.

2) The data of pigs under general anesthesia are used as the ground truth for quantitative analysis to avoid the influence of artifacts in human clinical data, because there is almost no physiological motion under general anesthesia. It is able to precisely demonstrate the DSA reconstruction effect for artifact removal and vascular structure preservation, without the motion contaminating ground truth.

\subsubsection{Overall Performance}
Quantitatively as the last column in Table \ref{tab1}, our proposed method achieves excellent performance with high PSNR of $43.19dB$ and high SSIM of $98.8\%$. Besides, visually as the last column in Fig.~\ref{exp}, it {gains the best} DSA imaging to precisely preserve the vascular structure and remarkably remove the motion artifact and other interference.

\subsubsection{Performance on Clinical Human Dataset}
As shown in Fig.~\ref{exp}, our proposed method achieves advanced DSA imaging with remarkable artifacts removal and precise vascular structures extraction, so that enables explicitly visual anatomy for clinical. And the other existing methods causes numerous blur artifacts in imaging, due to the difficulty of distinguishing between blood vessels and bone structures, by solely getting information from the fill images and completely dependent on the learning labels inherently with artifacts. Our method deals with such ill-posedness by 
learning the background equivalence between mask image and live image, and constraining the vascular consistency among motions, for non-contrast agent structure suppression and vascular enhancement. 
It enables clear background and precise vascular details even in complex areas with low contrast, significantly improving the visibility of blood vessels and providing strong support for clinicians to accurately diagnose, as shown in Fig.~\ref{exp} (e).

Specifically as enlarged regions in Fig.~\ref{exp}, our method further gains superiority for structural details imaging in complex region, indicated by circles, on the uAngio 960 dataset.
As shown in Fig.~\ref{exp}(a1)-(a5) enlarged regions, the blood vessel in the fill image is still difficult to clearly observe even under partial enlargement.
It demonstrates the necessity of improving the visibility of blood vessels in DSA imaging. Furthermore, it also illustrates the limitations of the solely fill image input paradigm especially in complex regions.
As enlarged regions in Fig.~\ref{exp} (a1)-(d1), because the blood vessel is covered by the complex bone structure in fill image, lots of artifacts and severe blood vessel loss occur in the results based on the synthesis methods, impeding the actually clinical value. 
And the enlarged regions in Fig.~\ref{exp} (a2)-(d2) show that oxygen tubes during interventional surgery are mistakenly identified as blood vessels imaging (indicated by arrow) in currently existing DSA methods, since both the similar structures as arc-shaped strips.  
For the area near the cochlea in Fig.~\ref{exp} (a3)-(d3), although there is no vascular structure, the compared DSA methods still seriously results in blurred patchy artifacts from the skeletal background. And our method in Fig.~\ref{exp} (e3) reliably suppresses all of these artifacts even in such cochlea region with complex bone structure distribution. 
As shown in Fig.~\ref{exp} (a4)-(d4) \& (a5)-(d5), brow bone and tooth in fill image mistakenly get dramatically contrast imaging (indicated by arrow) in compared methods, while vascular structures are massively missing.   
Remarkably, as shown in Fig.~\ref{exp} (e1)-(e5), despite facing the challenges of complex structures such as teeth, bones, and oxygen tubes, the proposed method still effectively enables precise vascular extraction even for tiny and end structure, and suppressed the generation of artifacts and maintained the clarity of the background, beneficial from Learning-based Subtraction Mapping Paradigm, Vascular Consistency Strategy, RDB-based Network Structure, and Mixup-based Data Self-evolution Strategy. 

\begin{figure*}[!htb]
    \centering
    \resizebox{0.7\textwidth}{!}
    {\includegraphics{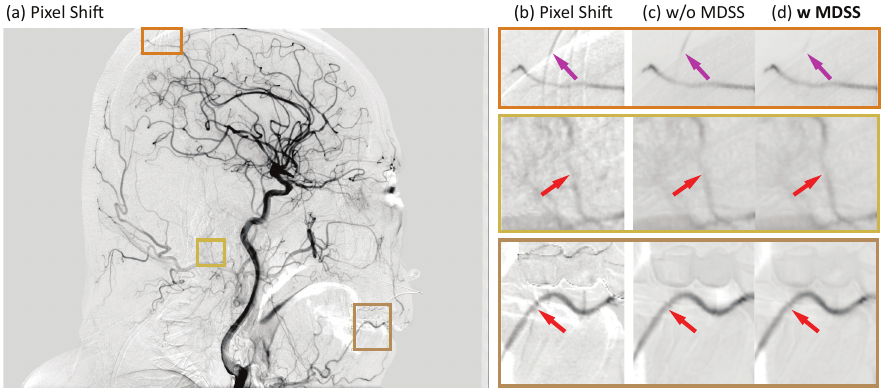}}
    \caption{
        Ablation performance of Mixup-based Data Self-evolution Strategy in qualitative experiments, and comparison with pixel shift.
        By comparing the effect of the MDSS, it can be observed that the MDSS strategy (d) not only enhances the visibility of the blood vessels in the second row, but also excludes the oxygen tube similar to the blood vessels in the first and third rows in (b)\&(c) and suppresses the teeth artifacts.
        }
    \label{figMDSS}
\end{figure*}

{Besides, as shown in Fig.~\ref{figAddition}, the proposed VCC-DSA also shows robust cross-dataset generalizability on the dataset with device variations (Philips Azurion 7 B20), achieving the precise DSA imaging with motion artifact suppressed and vessel structure preserved. Specifically, compared with the commercial algorithm in equipment, the proposed effectively remove the serious motion artifact, as shown in Fig. ~\ref{figAddition} (a1) vs. (a2) and (c1) vs. (c2) with blue indicted. These serious artifacts are caused by substantial motion of the head and neck, even resulting in compromised visualization of the vessel. And the VCS designed in VCC-DSA is able to explore consistency in motion, thereby extracting the vessel from artifacts. As shown in Fig.~\ref{figAddition} (a1) vs. (a2) and (c1) vs. (c2) with red indicted, as well as comparison in Fig.~\ref{figAddition}(b1)-(b2), the interference like brow ridge, breathing tube, bone and etc, also cause confusing structures in vessel DSA imaging. Beneficial from the MDSS designed in VCC-DSA, the interpenetration of blood vessels has evolved throughout training, enriching the vascular structure through self-updated datasets. As can be seen, the proposed VCC-DSA is designed following the intrinsic correlation before and after contrast agent injection, thus focusing on the consistency of vessel in any live and mask image pairs instead of specific dataset, as well as vessel structure learning, so that prompts the robust cross-dataset generalizability.}

\subsubsection{Performance on Animal Experimental dataset}

Quantitatively, as results in Table \ref{tab1}, the proposed method achieves significant improvement with PSNR and SSIM increased by 18.28dB and 7.786\%, respectively. Such performance superiority arises from effectively addressing ill-posedness and data dependency issues. The designed Learning-based Subtraction Mapping Paradigm adaptively aligns the background in mask \& live images, and distinguishably perceives the vascular imaging injected with contrast-agent. 
And the proposed Vascular Consistency Strategy and Mixup-based Data Self-evolution Strategy further enable the model intrinsically understand the vascular structure by results-inter structural-constraints and data-intra self-enhancement in training loop, removing data and label dependency. 
The RDB-based Network Structure promotes facilitates vascular structure extraction with more details.

Besides, as the last column in Table \ref{tab1}, the proposed method gains a fast inference time of $0.082$ second. It demonstrates that our method comprehensively gets superiority both on performance and efficiency.

{
As an external testing, this animal experimental dataset is directly evaluated using models trained on human-data for additional case-level analysis, clarifying the application scope of different scenarios. As shown in Fig.\ref{FigAnimal}, our method still demonstrates significant DSA imaging superiority with precise vascular imaging and robust artifact suppression, even with cross-species training and testing. Specifically, along with the contrast-agent flows as from Fig.\ref{FigAnimal}(a1) to (a3), our method (Figs.\ref{FigAnimal}(e1)-(e3)) consistently achieves vascular imaging close to the ground truth, while the comparative synthetic methods result in breakage of major vessels (as indicated by arrows in magnified blue regions), loss of tiny and terminal vessels (shown by arrows in magnified red regions), and generation of serious artifacts (highlighted by yellow circles). This superiority stems from: Learning-based Subtraction Mapping Paradigm that aligns background features between mask and live images while preserving contrast-enhanced vessels; Vascular Consistency Strategy and Mixup-based Data Self-evolution Strategy enables structural understanding with minimal data and label dependency; and RDB-based Network Structure that enhances detailed vascular feature extraction.
}

\begin{table}[]
                \vspace{-0.06in}
    \caption{
        ABLATION PERFORMANCES OF THE LEARNING-BASED SUBTRACTION MAPPING PARADIGM:
        The method with LSMP is significantly better than the method without MDSS in quantitative metrics.
    }
            \vspace{-0.05in}
            \small
    \setlength\tabcolsep{10pt}
    \label{tab2}
    \begin{center}
        \begin{tabular}{ccccc}
            \hline \hline
                method     & SSIM($\%$)$\uparrow$  & PSNR($dB$)$\uparrow$  \\ \hline
            w/o LSMP & 90.726 & 23.67680     \\
            \textbf{w LSMP (proposed)} &  \textbf{ 96.841} &     \textbf{32.61181}  \\ \hline \hline

        \end{tabular}
    \end{center}
\end{table}

\subsubsection{Ablation Study}

\textbf{Effectiveness of Learning-based Subtraction Mapping Paradigm.}
To demonstrate the effectiveness of Learning-based Subtraction Mapping Paradigm with mask image as background input, we conducted the ablation experiment by controlling the input of the background information and degrading to only live image.

As illustrated in Table~\ref{tab2}, the inclusion of learnable subtraction with mask image resulted in a 6.74\% increase in SSIM and a 37.7\% increase in PSNR.
This improvement is attributed to the fact that our Learning-based Subtraction Mapping Paradigm solves the ill-posed problem.

Visually in Fig.~\ref{mask}(d), Learning-based Subtraction Mapping Paradigm accurately extracts blood vascular (identified by red arrow), as well as successfully distinguish and remove other object similar to the vascular curve (identified by blue arrow), even for the area overlapped with the skull.
In contrast, fake vascular imaging (identified by blue arrow) occurs in Fig.~\ref{mask}(c) without using LSMP, because of failing to address the similarity between skeletal structure and blood vascular in these areas. And without using LSMP also causes the vascular lost in these areas that contrast of blood vessels in is low and overlaps with the skeletal structure, as identified by red arrow in Fig.~\ref{mask}(c). In addition, the inserted catheter is also incorrectly recognized as having a gray level similar to that of the blood vessel, and the skull edge is incorrectly recognized as having a shape similar to that of the internal carotid artery.

\begin{figure*}
    \centering
    \resizebox{0.96\textwidth}{!}
    {\includegraphics{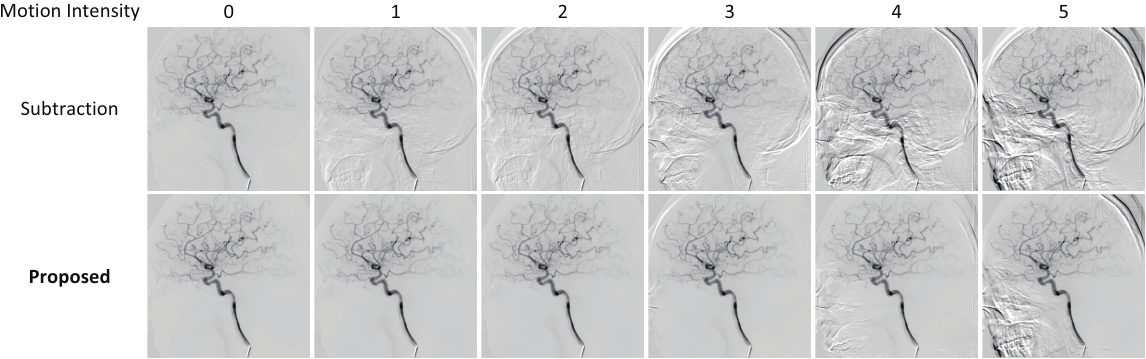}}
    \caption{
 The proposed method enables high-quality DSA imaging with precise vascular structure and robust artifact suppression for most common clinical motion conditions.
        }
                    \vspace{-0.05in}
    \label{FigMotion}
\end{figure*}

\textbf{Effectiveness of Vascular Consistency Strategy.}
To comprehensively evaluate Vascular Consistency Strategy, we analyze the influence of removing this strategy and its impacts with varying weight degrees.
We set $\lambda=0$ to degrade the model to the situation without VCS, and further adjust such consistency constraint parameter $\lambda$ with different value to reveal its effect for artifact removal in DSA imaging.

Fig.~\ref{fig4} shows that the increase in $\lambda$ further improves performance of the DSA imaging of our model. When $\lambda=0.8$, the quantitative indicators SSIM and PSNR are improved by 1.9\% and 25.9\% with the best performance.
At $\lambda=0.9$, the performance shows decline.

Visually in Fig.~\ref{lambda}, the reduction in artifacts is very significant with the increase in $\lambda$, especially after $\lambda$ exceeds 0.5. When $\lambda$ is 0.9, there was a slight loss of blood vessels at the end.
This indicates that the consistency strategy can effectively reduce artifacts when $\lambda$ is increased, but when the $\lambda$ value is too large, the loss of vascular information will occur due to the too low weight of the fidelity term in loss function Eq.\eqref{loss}.
And $\lambda=0.85$ is chosen as the best setting for VCS to enhance the vascular details and artifacts removal.

Particularly in Fig.~\ref{lambda}, we also make comparison with the Pixel Shift algorithm (registration-based method) of the commercial software in DSA imaging of United Imaging Healthcare. Our method shows significant superiority and achieves remarkable vascular imaging which is even covered with artifact in the commercial software, revealing its excellent DSA imaging and robust artifact suppression.

\begin{table}[]
    \caption{
        ABLATION PERFORMANCES OF MIXUP-BASED DATA SELF-EVOLUTION STRATEGY:
        The method with MDSS is superior to the method without MDSS in quantitative metrics.
        }
        \small
            \vspace{-0.05in}
    \begin{center}
        \label{MDSS}
        \begin{tabular}{ccc}
            \hline \hline
            & SSIM($\%$)$\uparrow$  & PSNR($dB$)$\uparrow$   \\ \hline
            w/o MDSS & 98.279 & 41.32683   \\
           \textbf{ w MDSS(proposed)}  &  \textbf{ 98.802} & \textbf{43.19122}   \\ \hline \hline

        \end{tabular}
    \end{center}
                \vspace{-0.13in}
\end{table}

\textbf{Effectiveness of the Mixup-based Data Self-evolution Strategy.}
To assess MDSS, we conducted an ablation experiment with the situation degraded to no-MDSS.  
As quantitative results in Table~\ref{MDSS}, the proposed MDSS outperforms the degraded situation with SSIM and PSNR increased by0.53\% and 4.51\%, respectively. The performance improvement reveal that using MDSS is good at optimizing data and enhancing model understanding on vascular structure.

As the visual results in Fig.~\ref{figMDSS}, the MDSS enables precise distinguishment between the blood vessels and the oxygen tube which ave similar structure, so that significantly promotes unambiguous and clear DSA imaging. It enhances the coherent and clear vascular structure (identified by red arrow), as well as eliminates the fake vascular imaging (identified by blue arrow) of oxygen tube edge caused by no-MDSS. Even with sparse vascular spatial proportion, the MDSS improves the model perception on intrinsic characteristics and distribution of vascular, and reduces the influence of the inherent artifact existing in label (as artifact in results of Pixel Shift Method), beneficial to its data-intra self-enhancement.

\subsubsection{Performance under Different Intensities of Motion Artifacts}

To further evaluate the robustness of our method, we test the performance under different intensities of motion artifacts. As shown in Table.~\ref{Motion}, our method achieves excellent PSNR and SSIM of $33.526dB$ and $93.486\%$ averagely on six motion intensities, and even for level 5 of large motion, still gains high PSNR and SSIM of $31.614dB$ and $92.774\%$, revealing its strong robustness for motion artifacts. The motion intensity is divided into 6 levels, and is simulated on the carefully selected clinical human data with small motion as much as possible. 

Detailed as shown in Fig.~\ref{FigMotion}, our method is robust to deal with different intensities of motion artifacts for precise and high-definition DSA imaging. Visually for the Levels 1-2 of common motion intensity in clinical and the Level 0 of no-motion, our method effectively maintain the clarity of vascular details and suppresses all artifacts. Even under serious motion conditions with Levels 3-4 which are rare in clinical, it is still good at preserving the details of blood vessels and eliminating most of the artifacts. And for extreme motion of Level 5, the vascular structure especially including tiny blood vessels is still successfully extracted with coherent and clear imaging for clinicians, though a few facial skeletal motion artifacts exists.
It can be seen that the proposed method enables high-quality DSA imaging with precise vascular structure and robust artifact suppression for most common clinical motion conditions.

\begin{table}[]
    \caption{
QUANTITATIVE RESULTS UNDER DIFFERENT INTENSITIES OF MOTION ARTIFACTS 
        }
    \setlength\tabcolsep{3pt}
    \small
    \begin{center}
        \label{Motion}
        \begin{tabular}{ccccccc}
            \hline \hline
          Motion Intensity  & 0 & 1 & 2 & 3 & 4 & 5 \\ \hline
            SSIM($\%$)$\uparrow$  & 94.273 & 94.011 & 93.660 & 93.253 & 92.943 & 92.774   \\
            PSNR($dB$)$\uparrow$  & 34.935 & 34.680 & 34.121 & 33.339 & 32.468 & 31.614   \\ \hline \hline

        \end{tabular}
    \end{center}
\end{table}

\begin{figure*}[!htb]
    \centering
    \resizebox{0.7\textwidth}{!}
    {\includegraphics{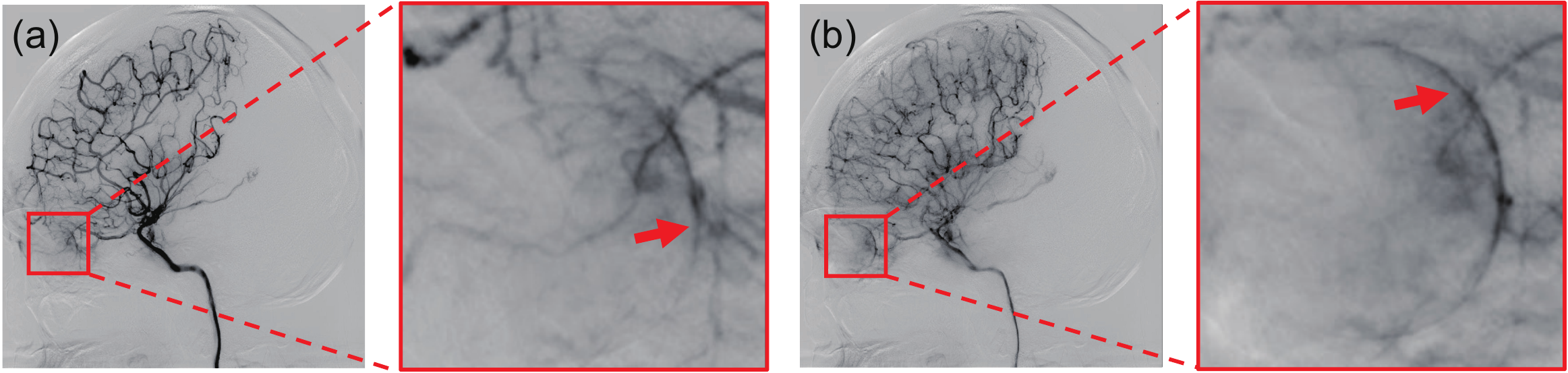}}
    \caption{
Failed cases in experiments.
        }
                    \vspace{-0.05in}
    \label{lim}
\end{figure*}

{\subsubsection{Failed case analysis}
The failed cases of the proposed method are exhibited in Fig.~\ref{lim}. As red arrow indicated in the enlarged region, partial artifact from the eyeball boundary is maintained. As indicated by the red arrows in the enlarged region, partial artifacts along the eyeball boundary are maintained. After data inspection, two key factors were identified: first, this patient’s eyeball exhibited active and continuous rapid rotational motion during intraoperative DSA imaging. Second, this active motion induced dynamic displacement of the surrounding vascular structures through mechanical traction. Consequently, this activate eyeball motion is unpredictable and difficult to extract, while its coordinated motion with peripheral vessels further complicates its isolation, especially the curvature similar to blood vessels as the red arrow indicated in Fig.~\ref{lim}.}

\section{CONCLUSION}

In this paper, we propose a novel VCC-DSA model for the precise DSA imaging with vascular structure enhancement and motion artifact suppression.
It is beneficial from:
1) a Learning-based Subtraction Mapping Paradigm to solve the ill-posed problem with solely live image and enable the stable DSA imaging result;
{2) a network based on RDB and detail enhancement design according to the fine and rich characteristics of the vascular structure, so that improves the performance of the network in DSA images;}
{3) a Vascular Consistency Strategy to extract intrinsically consistency from the various relative motions in mask-live images, so that spontaneously
distils the vascular structure with contrast-agent development and robustly suppress motion artifacts;}
4) a Mixup-based Data Self-evolution Strategy for data-intra self-enhancement in training loop to optimize training data and promote model better learning the vascular features.
Our framework does not rely on data screening and data labeling, which makes the model more practical.
A large number of experiments demonstrate the effectiveness and robustness of our model in DSA imaging.
The image quality advantage of our method can reduce the artifacts caused by patient movement in clinical applications, thereby reducing the possibility of re-acquiring DSA data, saving valuable time for the treatment of cardiovascular diseases, especially acute stroke.

\section*{Declaration of competing interest}
The authors declare that they have no known competing financial interests or personal relationships that could have appeared to influence the work reported in this paper.

\section*{Acknowledgments}
This work was supported in part by the State Key Project of Research and Development Plan under Grant 2022YFC2408500, in part by the National Natural Science Foundation of China under Grant T2225025, 62561160154 and 62101249, in part by the Key Research and Development Programs in Jiangsu Province of China under Grants BE2021703 and BE2022768, in part by Jiangsu Province Science Foundation for Youths under Grant BK20220825, in part by the ``Shuangchuang'' Doctor program of Jiangsu Province under Grant JSSCB20220202, in part by the China Postdoctoral Science Foundation under Grant 2021TQ0149 and 2022M721611,
in part by the Interdisciplinary Research Program for Young Scholars under Grant 2024FGC1004 from Southeast University, China and the Fundamental Research Funds for the Central Universities, China under Grant 2242025F10004. This research work is supported by the Big Data Computing Center of Southeast University, China.
 
\bibliographystyle{model2-names.bst}\biboptions{authoryear}
\bibliography{reference}

\end{document}